\documentclass[12pt]{article}
\usepackage{geometry}
\usepackage[backend=biber,style=nature,citestyle=nature,doi=false]{biblatex}
\usepackage{authblk}
\usepackage{physics}
\usepackage{dsfont}
\usepackage{subcaption}
\usepackage{graphicx}
\usepackage{amsmath}
\usepackage{comment}
\usepackage{xcolor}
\usepackage[normalem]{ulem}
\usepackage{mathtools}

\DeclareSourcemap{
  \maps[datatype=bibtex]{
    \map{
      \step[fieldset=url, null]
      \step[fieldset=issn, null]
      \step[fieldset=month, null]
      \step[fieldset=eprint, null]
            \step[fieldset=isbn, null]
    }
  }
}

\DeclareSourcemap{
  \maps[datatype=bibtex]{
    \map{
      \step[fieldset=issn, null]
    }
  }
}

\title{Quantum Reservoir Computing Implementations for Classical and Quantum Problems}

\author[1]{Adam Burgess\thanks{a.burgess@hw.ac.uk}}
\author[2]{Marian Florescu\thanks{m.florescu@soton.ac.uk}}

\affil[1]{SUPA, Institute of Photonics and Quantum Sciences, Heriot-Watt University, Edinburgh, EH14 4AS, UK}
\affil[2]{Optoelectronics Research Centre, University of Southampton, Southampton  SO17 1BJ, United Kingdom}

\date{}
 
\begin{document}
 \maketitle
 \begin{abstract}
Quantum reservoir computing has emerged as a promising paradigm within the field of quantum machine learning, harnessing the inherent properties of quantum systems to optimise and enhance information processing capabilities.

Here, we explore the potential of quantum-inspired machine learning methodologies by leveraging the complex dynamics of quantum reservoirs to address computationally challenging tasks with enhanced efficiency and accuracy. To this end, we employ an open quantum system model comprising two-level atomic ensembles coupled to Lorentzian photonic cavities to construct a quantum physical reservoir computer layer for a recurrent neural network. We evaluate the effectiveness of this approach by applying it to a standard machine learning image-recognition problem and benchmarking its performance against a conventional neural network of similar architecture, but lacking the quantum physical reservoir computer layer. Remarkably, as the dataset size increases, the quantum physical reservoir computer outperforms the conventional neural network,  requiring fewer training epochs and a smaller dataset to achieve comparable accuracy. Furthermore, we employ the quantum physical reservoir computing approach to model the dynamics of open quantum systems, focusing on atomic system ensembles interacting with a structured photonic reservoir associated with a photonic band-gap material. Our results reveal that the quantum reservoir computer provides equally powerful representations for quantum dynamical problems, maintaining effectiveness even under constraints of limited training data.
\end{abstract}

\begin{refsection}
\section*{Introduction}\label{sec1}

Many physical systems exhibit evolution that cannot be described by closed-form solutions. Conventional strategies for addressing this problem often rely on approximation schemes, such as the Markovian approximation, which assumes a clear separation between the environmental and system dynamical timescales, or perturbation approaches such as Bloch-Redfield master equations, which are second order in coupling strengths~\cite{TheoryOQSBook} and the cumulant expansion approach that accounts for additional system correlations beyond the mean field ~\cite{Plankensteiner_2022}. Although numerically exact methods have also been developed to tackle the challenges emerging in many-body physics, their applicability is often hindered by unfavourable scaling with the system size and complexity. Moreover, it is precisely for these large and highly complex quantum systems that approximate methods break down, and numerical approaches are equally ineffective in overcoming the underlying challenges.

Recent progress in understanding the mathematical foundations of artificial neural networks~\cite{LeCun2015} has driven a growing interest in the popularity and feasibility of data-driven approaches to solving physical problems, capitalising on the intrinsic statistical properties of many physical systems. This is further driven by the efficient implementation of neural network techniques, which rely predominantly on cost-effective linear transformations. However, the primary computational cost lies in the training process through back-propagation in which the internal weights of the network are being optimised. This cost is balanced by the fact that once these networks are trained, they exhibit remarkable efficiency. In particular, neural networks have shown connections to conventional physics methods such as the renormalisation group ~\cite{RenromToDeepLearn} and tensor networks ~\cite{BoltzmannTensor, Levine_2019, DLQE}.  The effectiveness of these approaches has been demonstrated in solving various challenging problems in physics, including the implementation of machine learning in quantum computing \cite{Biamonte2017, doi:10.1098/rspa.2017.0551, Dunjko_2018}, quantum many-body problems \cite{NNManyBody, MLManyBody}, characterisation of phase transitions in complex condensed matter systems \cite{vanNieuwenburg2017, Carrasquilla2017}, and quantum state tomography \cite{RNNNonMarkov, Torlai2018}. Furthermore, optimisation strategies developed for machine learning have found applications in quantum circuit design \cite{TrainQDL, RLinQC} and in the prediction of the dynamics of quantum many-body systems \cite{MLManyBody, RNNNonMarkov}.  

\begin{figure*}[ht]
     \centering
      \begin{subfigure}[b]{0.33\textwidth}
         \centering
         \includegraphics[width=\textwidth]{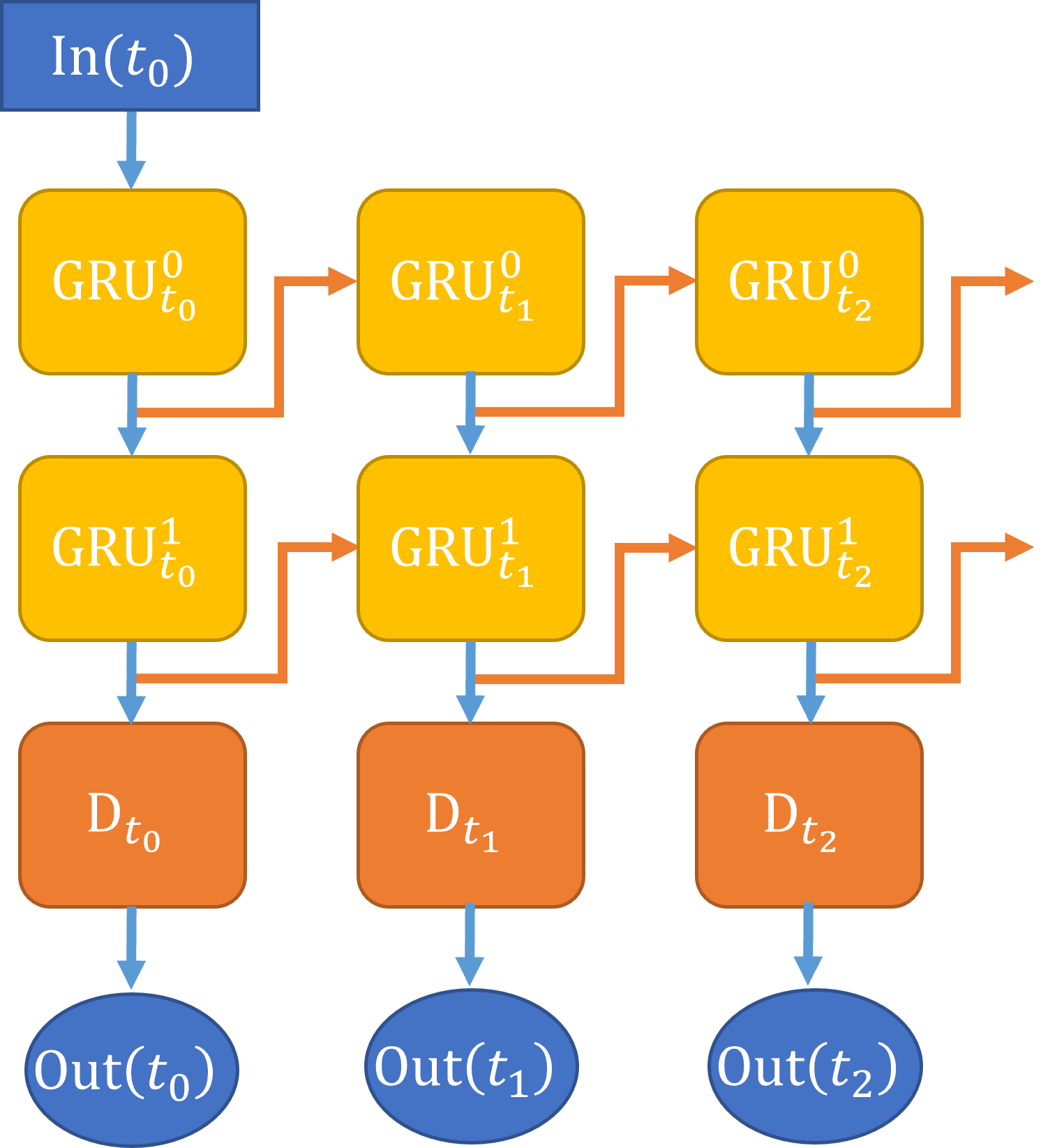}
         \caption{}
         \label{fig:RNNarch}
     \end{subfigure}
      \begin{subfigure}[b]{0.6\textwidth}
         \centering
         \includegraphics[width=\textwidth]{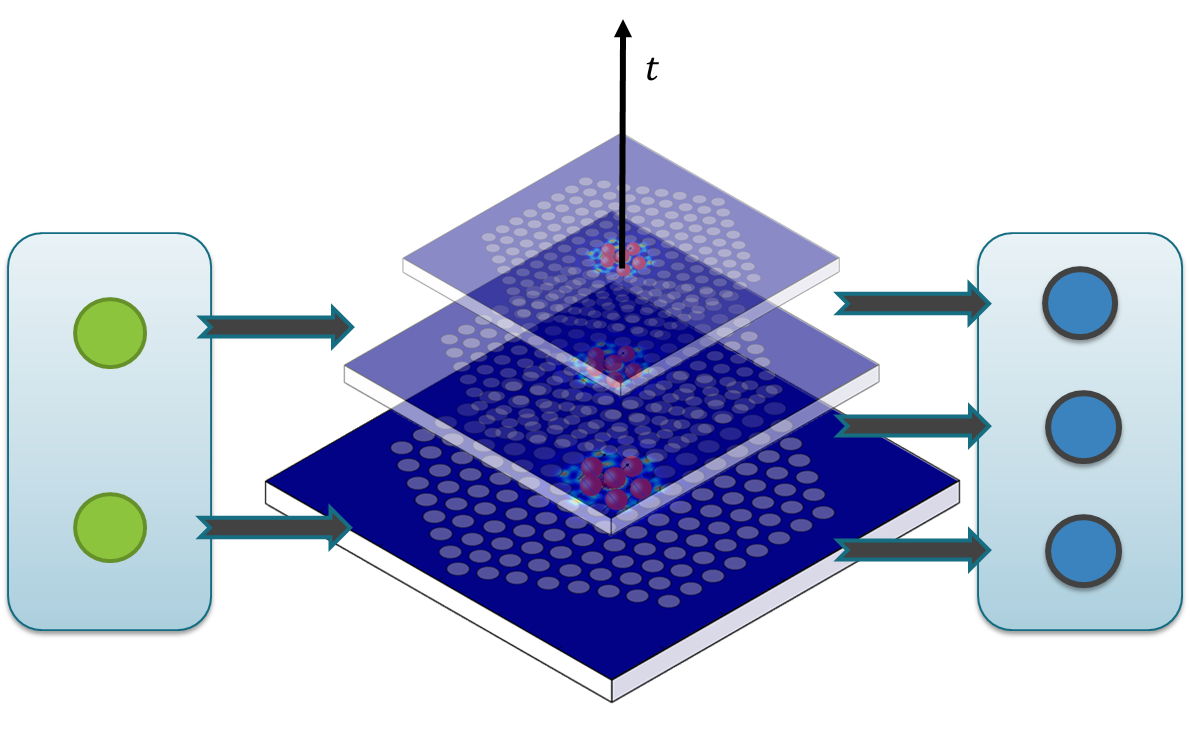}
         \caption{}
         \label{fig:QRes}
     \end{subfigure}
     \hfill
        \caption{(a) Schematic of a recurrent neural network, showing the input to the output pathways including  GRU layers~\cite{DBLP:journals/corr/ChungGCB14}  (a type of RNN layer). Information is passed vertically through layers (blue arrows) and horizontally (orange arrows) between GRU timesteps to embed memory of previous timesteps. In such an RNN, many internal weights of the GRU layer need to be updated by back-propagation, making it an effective but computationally expensive architecture. (b) Schematic of the Quantum Physical Reservoir Computing network, an input is fed into the open quantum system and the time series of the quantum state evolution is fed into the dense layer at the end of the network (here, only the dense layer weights need to be updated.}
        \label{fig:Intro}
\end{figure*}

However, modelling the dynamics of open quantum quantum systems poses even greater challenges compared to the conventional closed quantum systems. For open quantum systems, neural network models must be able to account for complex interactions with an environment which may posses inaccessible microscopic degrees of freedom. The task becomes more difficult for systems that exhibit strong non-Markovian behaviour, where temporal correlations can occur over a multitude of timescales and, as such, significantly increase the necessary complexity of any scheme for resolving the dynamics. Consequently, even simulating low-dimensional open quantum systems can be computationally expensive and impractical. Machine learning models offer potential solutions for modelling the dynamics of such complex systems. Neural networks can effectively represent and capture interactions between the system of interest and complex environments without making assumptions about their nature. Recurrent neural networks that are able to account for memory of previous states have also been shown to efficiently model non-Markovian dynamics \cite{RNNNonMarkov, GRU} too (see Fig.~\ref{fig:RNNarch} for a schematic of the recurrent network). 

\begin{figure}[h]
    \centering
    \includegraphics[width=0.75\textwidth]{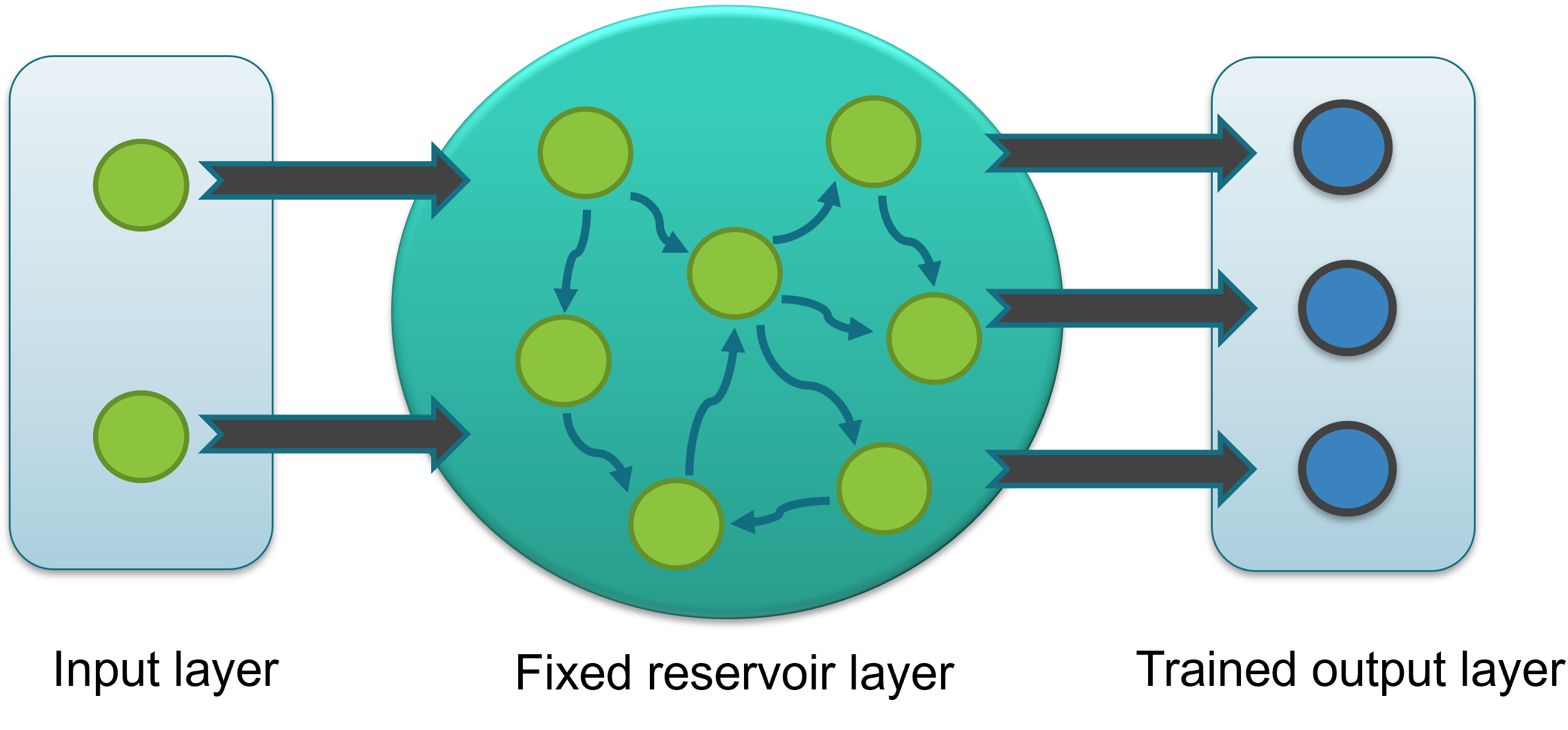}
    \caption{A schematic of a conventional reservoir computing architecture. Inputs are fed into a large reservoir layer, then readouts are taken into an output layer and only weights in the output layer are optimised. }
    \label{fig:RC}
\end{figure}

Recently, reservoir computing has emerged as a prominent paradigm for developing recurrent neural networks and has been actively explored~\cite{LUKOSEVICIUS2009127,ESM,LSM}. 
Reservoir computing functions on the principle of mapping inputs into high-dimensional computational spaces, called `the reservoir', then outputting information from the reservoir into a reduced neural network. The training process involves only the reduced output network, eliminating the need to update the internal reservoir parameters. This greatly reduces the computational cost associated with backpropagation.  A pictorial schematic of this is shown in Fig.\ref{fig:RC}.   The computational efficiency achieved by reservoir computing is crucial, considering that high-performance computers and distributed systems are often required for calculations in many successful machine-learning projects, such as natural language processing \cite{NLP}  and machine-learning systems capable of playing strategy games such as DeepMind's AlphaGo \cite{Silver2016}.  Therefore, reducing computational overhead is not only preferable, but necessary \cite{CostNN}. Furthermore, since the reservoir weights remain unchanged, the reservoir can be used for multitasking by feeding its outputs into multiple independent neural networks \cite{MultitaskingRC}. Traditionally, reservoirs are large, unaltered neural networks with random initialisation of their weights. However, recent developments have explored the capability of physical systems as reservoir computers, wherein the system's dynamical evolution allows for the mapping of input parameters, defining initial conditions to a high-dimensional computational space, or the phase space of the dynamical system. By performing readouts on this dynamical system, information can be fed into the reduced neural network. This concept is known as ``physical reservoir computing" \cite{Nakajima_2020}. It has shown promise as computations can be offloaded to purely dynamical systems, and various implementations have been proposed, ranging from soft robotics to spintronics systems~\cite{ResSpin1,ResSpin2,ResSoft}.   

In the context of large physical systems with strong interactions and numerous internal degrees of freedom, the dynamics often exhibit significant non-linearity \cite{Chang2014, TUSZYNSKI1989179, NonLinearQD}, which satisfies one of the requirements for a reservoir computer. If we consider a many-body quantum system as the physical reservoir, the entanglement among subsystems (capturing the non-trivial correlations encoded in the many-body wave function) provides access to a rich set of internal degrees of freedom. This makes quantum physical reservoir computers (QPRCs) a desirable prospect. However, when dealing with quantum mechanical systems, we need to account for their interaction with the environment to be able to perform readouts. As a result, QPRCs are no longer isolated systems, and an open quantum system framework is necessary to describe their behaviour, as depicted in Fig.~\ref{fig:QRes}. In the context of reservoir computing, this turns out to be a major advantage as the dynamics is now mapped onto an even larger computational space, combining the Hilbert spaces of both the system and the environment.

In this work, we exploit the dynamics of an open quantum system as a generic quantum physical reservoir to establish the reservoir computing framework. We do this by introducing the physical platform for implementing the proposed QPRC and demonstrating its effectiveness against both a typical pattern recognition problem within the domain of machine learning and
a fundamental open quantum mechanical problem involving quantum emitters coupled to a structured photonic reservoir.

\section*{Quantum Physical Reservoir System}
\label{PhysRevSys}

For a physical reservoir computer to be effective, it is crucial that the reservoir exhibits predictable behaviour and consistently produces the same output for a given input. One modality to achieve this is by utilising a dissipative system, for example, an atomic system coupled to a leaky cavity, which approaches an equilibrium state under well-controlled evolution. By lowering the system temperature, this equilibrium state aligns closely with the ground state of both the atomic systems and the local reservoir. Laser driving can be used to generate effective initial conditions based on a given input data set, ensuring a reliable and predictable system evolution \cite{ControlCavityQD, Miguel-Sanchez_2013, QDControl2}.

To effectively implement the QPRC approach, it is necessary to identify a suitable large-scale open quantum system model that can be efficiently sampled. In this study, we focus on many-body atomic systems confined within a single lossy photonic cavity \cite{burgessNatom, BurgessMemoryEffects} as the model system of interest. The cavities considered here can be realised in a variety of implementations, including Fabry-Perot cavities, photonic crystal cavities, plasmonic nanocavities, and whispering-gallery mode cavities~\cite{Yang2024, Nomura2010,Faraon2010,Birowosuto2014,Nozaki2012,Day2023, Pres2023,deOliveira2024}.

The dynamics of the composite system are governed by the Hamiltonian:
\begin{align}
    H = &\sum_{i=1}^N\omega_0\sigma_i^+\sigma_i^- + \sum_\lambda \omega_\lambda a^\dag_\lambda a_\lambda +  i\sum_{i,\lambda}g_\lambda(a_\lambda \sigma^+_i -a^\dag_\lambda \sigma^-_i ) + \sum_{i\neq j}\Lambda\sigma^+_i\sigma_j^-,
    \label{eqn:RWAH}
\end{align}
where $\sigma_i^+$and $\sigma_i^-$ are the excitation and de-excitation operators for the $i^\textrm{th}$ atomic system respectively, $a_{\lambda}$ and $a^{\dagger}_{\lambda}$ are the bosonic field annihilation and creation operators, $\omega_0$ and $\omega_{\lambda}$ are the atomic transition and the $\lambda$-boson mode frequencies,  $g_\lambda$ is the coupling strength of the atomic system and the $\lambda$-boson mode and $\Lambda$  is the dipole-dipole coupling strength between atoms~\cite{Pfeifer,AllenEberly,SuperRadiancePBG}. This is equivalent to a Dicke model~\cite{Dicke} for $N$ atoms with an interatom coupling. 
We have chosen a highly symmetric atomic system to generate time-resolved data efficiently. To this extent, we also limit the investigation to the single-excitation regime. We also assume an initial condition given by 
\begin{equation}
    \phi(0) = c_0\psi_0+\sum_i^N c_i(0)\psi_i +\sum_\lambda c_\lambda(0) \psi_\lambda,
\end{equation}
where $\psi_i = \ket{i}_A\ket{0}_B$ is the state in which the $i^\textrm{th}$ atom is in its excited state and all other atoms and the photonic cavity are in the ground state. $\psi_\lambda = \ket{0}_A\ket{\lambda}_B$ represents all atoms in their ground state, and the cavity system has its $\lambda$ mode excited. $\psi_0 = \ket{0}_A\ket{0}_B$ denotes the ground state of the entire system.

Then the time evolution of this initial state is given by
\begin{equation}
    \phi(t) = c_0\psi_0 + \sum_i^N c_i(t)\psi_i +\sum_\lambda c_\lambda(t) \psi_\lambda.
\end{equation}
Using an adiabatic cancellation of the photonic mode variables $c_\lambda$ 
and introducing the memory kernel 
\begin{equation}
 G(t) = \sum_\lambda g_\lambda^2 e^{i(\omega_0-\omega_\lambda)t},
\end{equation}
the equations of motion for the atomic degrees of freedom can be expressed as
\begin{align}
    \Dot{c}_i =& -i\Lambda(c_+-c_i) - \int^t_0G(t-t_1)c_+(t_1)dt_1,
    \label{eqn:ConvolutionRC}
\end{align}
where $c_+ =\sum_i^N c_i$.
The non-Markovian or memory effects in the system's dynamics are associated with this convolution over the memory kernel at previous times. This is precisely the aspect of the QPRC implementation that we intend to leverage.

The time evolution of this system is then determined by the distribution of the coupling strengths given by the spectral density
\begin{equation}
    J(\omega) = \sum g_\lambda^2 \delta(\omega-\omega_\lambda),
\end{equation}
which is related to the memory kernel by a simple relation
\begin{equation}
G(t) = \int d\omega J(\omega )e^{i(\omega_0-\omega)t}.
\end{equation}
For a lossy photonic cavity resonant with the atomic transitions, it is common to model the spectral density by a Lorentzian distribution of the form
\begin{equation}
    J(\omega) = \frac{\lambda\gamma}{\gamma^2+(\omega-\omega_0)^2},
\end{equation}
with $\lambda$ the coupling strength of the atoms to the cavity mode, and $\gamma$ determining the spectral width of the lossy cavity \cite{NMDJC,TheDJC,ControlJC}. 

As a result, the time evolution of the excited state amplitudes for the $i^\textrm{th}$atom is given by 
\begin{align}
c_i(t)&= (c_i(0) -\frac{c_+(0)}{N})e^{i\Lambda t} \nonumber+ \frac{c_+(0)}{N}e^{-\mu^* t}\left(\cosh(\Gamma t) + \frac{\mu}{\Gamma}\sinh(\Gamma t)  \right),
\label{eqn:cit}
\end{align}
where we have introduced the following for convenience the following notations 
\begin{align}
    \Lambda_{FC}=J(N-1), \,\,\Gamma =\frac{1}{2}\sqrt{ \gamma^2-2i\gamma \Lambda_{FC} -\Lambda_{FC}^2-4\lambda N}, \,\, 
    \mu = \frac{1}{2}\big(\gamma-i\Lambda_{FC}\big).
\end{align}

To employ the physical system as a QPRC layer, we need to establish a procedure for inputting data into the QPRC layer and subsequently outputting it to the neural network. Given a stream of input data denoted as $\mathcal{I}$, we define a bijection $f$ that maps $\mathcal{I}$ to the multi-valued ensemble of values $c = {c_i}$. Additionally, we require a mapping function $g$ that transfers the time-evolved wavefunction $\phi(t)$ to $\mathcal{O}$, representing the output of the QPRC layer. This output is then directly passed into the conventional neural network layers. In our approach, we utilize the real and imaginary components of the excited state amplitudes at different timesteps ${t_j}$ as the map $g$. Specifically, $g:\phi(t)\rightarrow {\Re{c_i(t_j)},\Im{c_i(t_j)}}$. This mapping can be likened to performing non-demolition measurements on the system or conducting quantum state tomography \cite{Cramer2010, Grangier1998}.

\section*{Quantum Physical Reservoir for Image Recognition}
\label{QPRCIm}

\begin{figure}[htp]
     \centering
     \begin{subfigure}{0.9\textwidth}
         \centering
         \includegraphics[width=\textwidth]{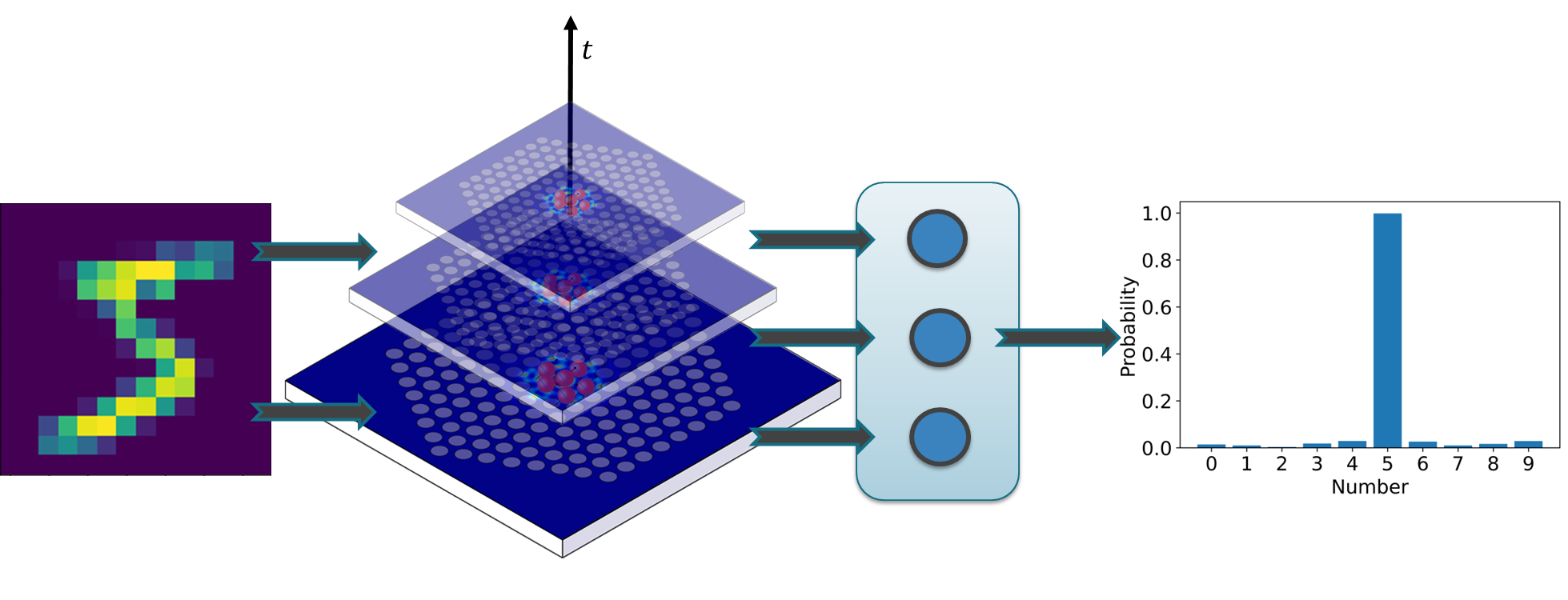}
         \caption{}
         \label{fig:Mnistarch}
     \end{subfigure}
     \\
     \begin{subfigure}{0.4\textwidth}
         \centering
         \includegraphics[width=\textwidth]{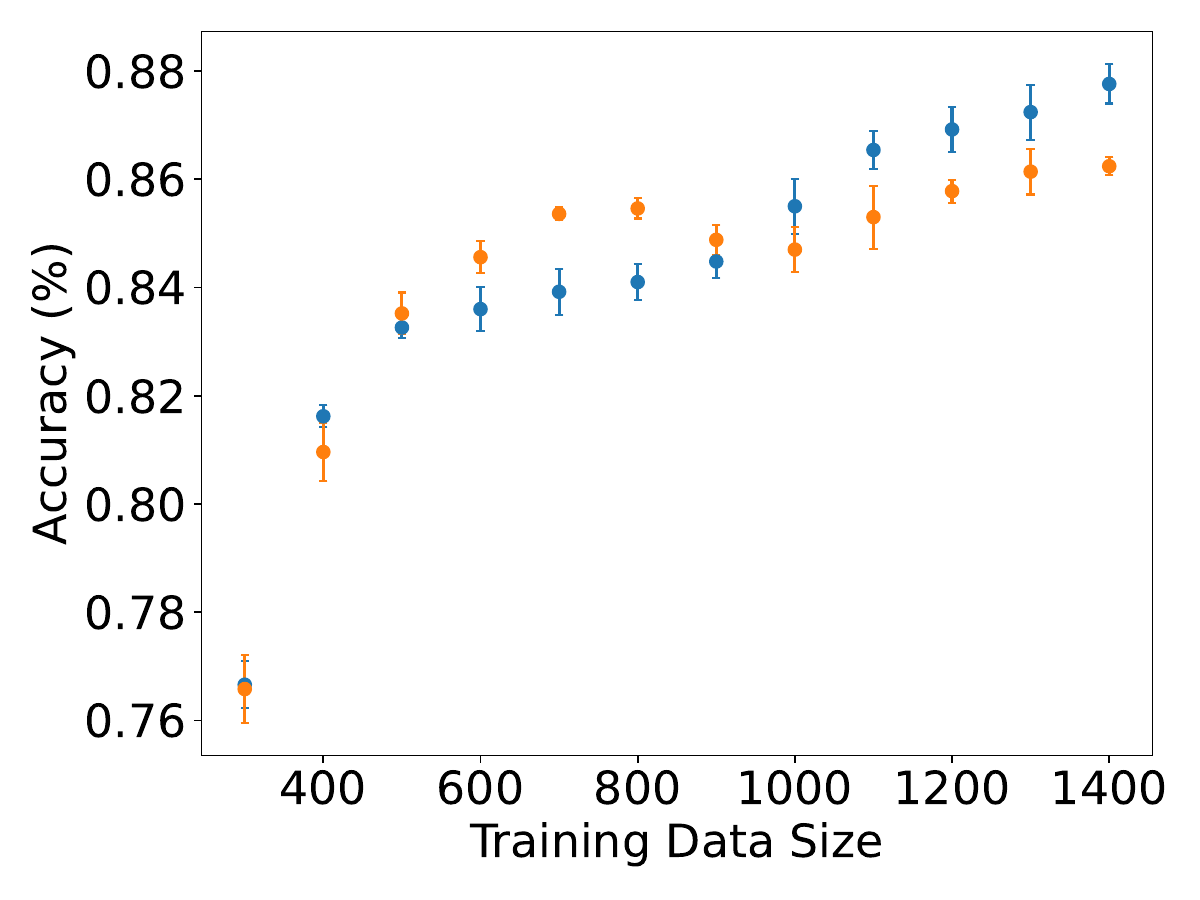}
         \caption{}
         \label{fig:ResSizeSmall}
     \end{subfigure}
     \begin{subfigure}{0.4\textwidth}
         \centering
         \includegraphics[width=\textwidth]{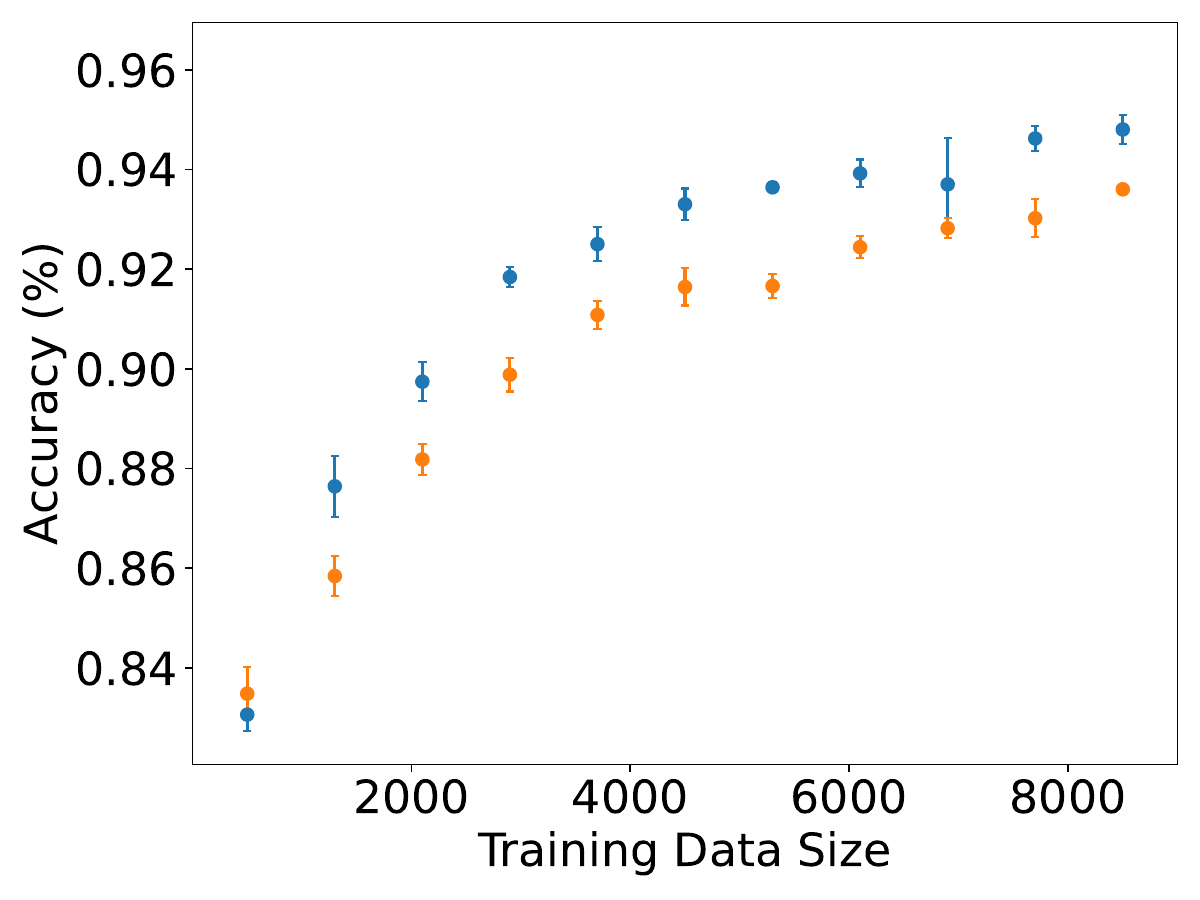}
         \caption{}
         \label{fig:ResSizeLarge}
     \end{subfigure}
    \begin{subfigure}{0.4\textwidth}
         \centering
         \includegraphics[width=\textwidth]{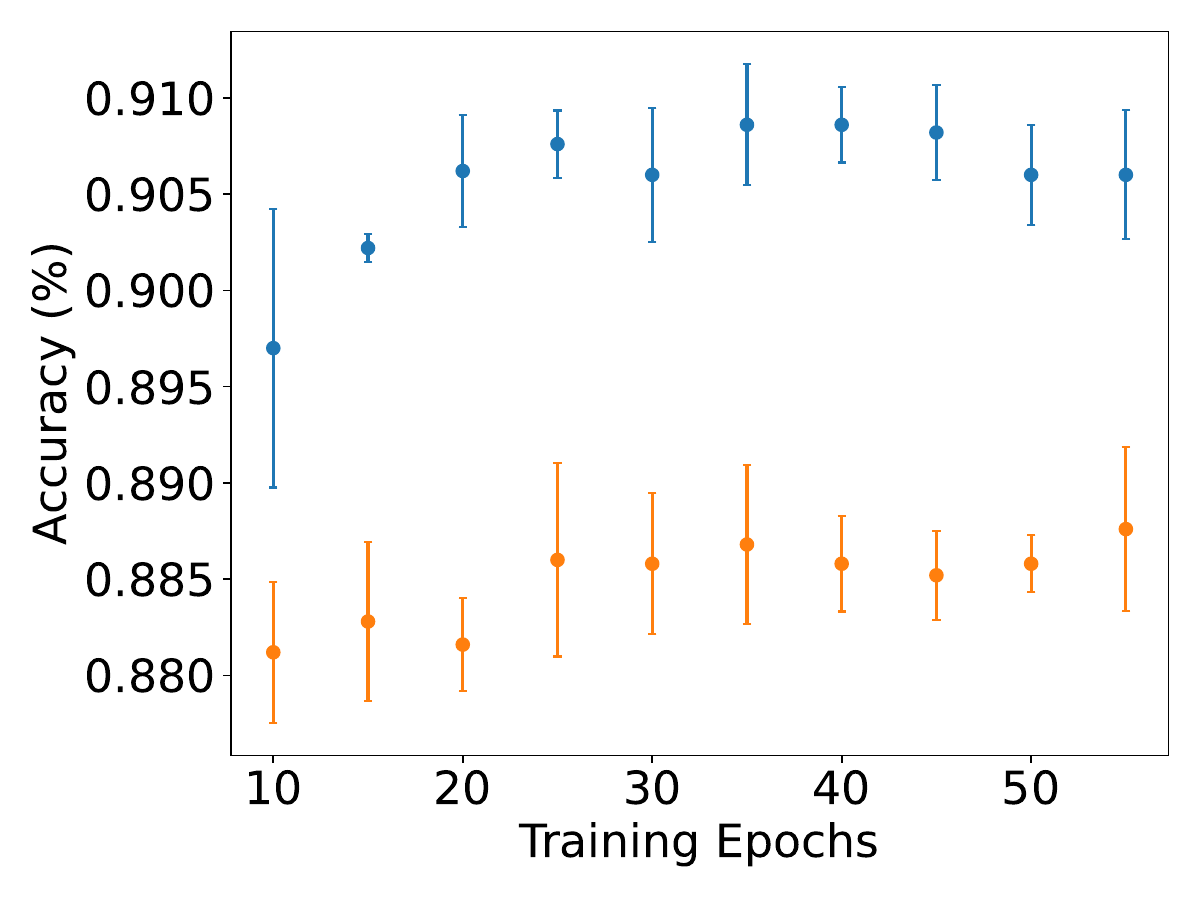}
         \caption{}
         \label{fig:ResEpoch}
     \end{subfigure}
        \caption{(a)  Schematic of the Quantum Physical Reservoir network used for  image recognition; pixel data is converted into an initial condition for the open quantum system, and the time series of the quantum state is fed into the dense layer at the end of the network. (b-d) The accuracy, as a fraction of correct predictions, of the QPRC approach (blue) and the conventional neural network without the QPRC layer (orange) for (b,c)  varying sizes of training data (d) varying numbers of training epochs.}
        \label{fig:MNISTS}
\end{figure}

We apply now the dynamical process of the QPRC framework to address a fundamental problem in machine learning, namely pattern recognition using the well-known MNIST database \cite{deng2012mnist}. This database comprises grayscale images of handwritten digits from 0 to 9, and the objective is to successfully classify each image according to the number being represented. To effectively leverage the reservoir computing approach, it is necessary to establish an isomorphism between the pixel data of each image and the initial condition of the system. To achieve this, we directly map the pixel brightness of the $2n$ and $2n+1$ pixels, where we convert the 2D image into a 1D vector, to the real and imaginary components of $c_n(0)$. Subsequently, we normalize the wavefunction $\phi(0)$. 

Mathematically, for the $i^\textrm{th}$ pixel in an image with brightness $P_{i}$, we define the mapping as $c_i = (P_{i} +iP_{i+1})/\mathcal{N}$, where $\mathcal{N}$ represents a normalization constant. Next, we evolve the wavefunction in time (as per Eqn.\ref{eqn:cit}) and sample it at $n=50$ equidistant time steps up to $n\cdot\omega_0^{-1}$. The resulting time series' real and imaginary components are then passed into a reduced neural network. This reduced neural network consists of a single dense layer with a dimension of $n_\textrm{nn}=128$. A schematic diagram illustrating this procedure is provided in Fig.\ref{fig:Mnistarch}. The advantage of this approach lies in the fact that we only need to perform gradient descent on the small final dense layer of the network, thereby simplifying the complex step of readjusting network weights.

To evaluate the performance of the QPRC network, we have conducted two numerical experiments. Firstly, we have compared the results generated using the QPRC network to those generated by a conventional neural network without the QPRC layer, using ensembles of training data of varying sizes. The results of this experiment are presented in Figs. \ref{fig:ResSizeSmall} and \ref{fig:ResSizeLarge}. We note that the QPRC neural network outperforms the conventional neural network for intermediate and large training data sizes (with a crossover point at around 900 training sets), achieving an accuracy of over 90\% in the prediction capabilities. This result is particularly significant given the increasing relevance of big data and the escalating computational costs associated with large neural networks used in natural language processing systems for which finding more efficient ways to handle large datasets is increasingly necessary. Moreover, our results suggest that the QPRC neural network generates a more effective representation of the studied dataset. To demonstrate this, we have compared the performance of the QPRC network to that of a conventional neural network with varying number of training epochs. The results presented in Fig.\ref{fig:ResEpoch} clearly demonstrate that the QPRC system consistently outperforms the non-QPRC network across all tested values of training epochs. Each of the numerical experiments depicted in Fig.\ref{fig:MNISTS} was repeated five times at each sampling point, and the error bars on the graph represent the standard deviations derived from these five samples. In these experiments, we utilized a single dense layer with a dimension of 128, employing a hyperbolic tangent ($\tanh$) activation function. The optimization was performed using the Adam optimizer with a learning rate of 0.001, and 30 training epochs were executed.

The QPRC platform we have introduced can be further optimized to enhance its performance and accuracy. For example, exploiting the correlations within the dataset can guide the selection of the QPRC system. Given the strong nearest-neighbour correlations exhibited by pixel data, a nearest-neighbour type of interaction in the quantum system could enhance the effectiveness of the approach. Such a system could consist of a 2D lattice of atomic systems embedded in an array of photonic cavities with controllable cavity-cavity positioning. This architecture, utilizing a 2D lattice geometry, would induce nearest-neighbour interactions between the atomic systems and generate commensurate nearest-neighbour correlations similar to the pixel data.

\section*{Quantum Physical Reservoir Computing Approach for Open Quantum Systems}
\label{QPRC2Q}

\begin{figure}[htp]
     \centering
     \begin{subfigure}{0.8\textwidth}
         \centering
         \includegraphics[width=\textwidth]{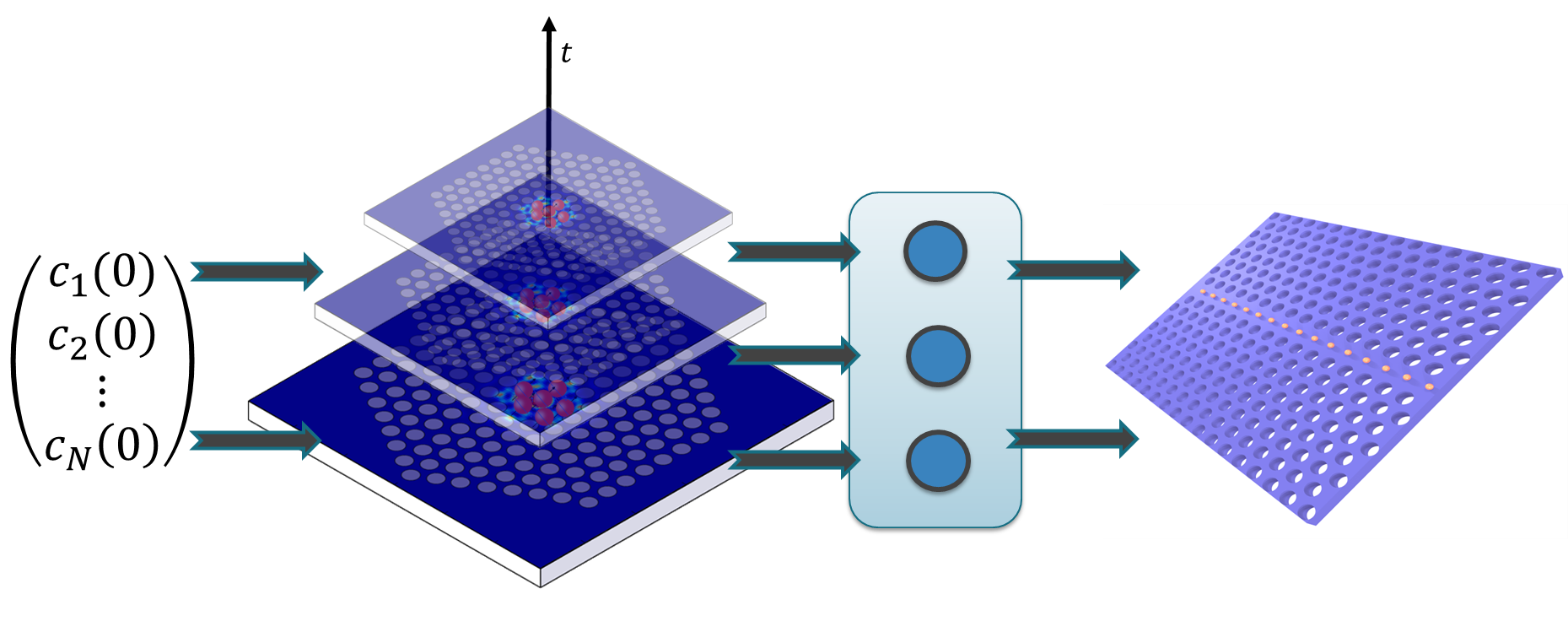}
         \caption{}
         \label{fig:Q2Qarch}
     \end{subfigure}
      \\
     \begin{subfigure}[b]{0.38\textwidth}
         \centering
         \includegraphics[height=0.75\textwidth]{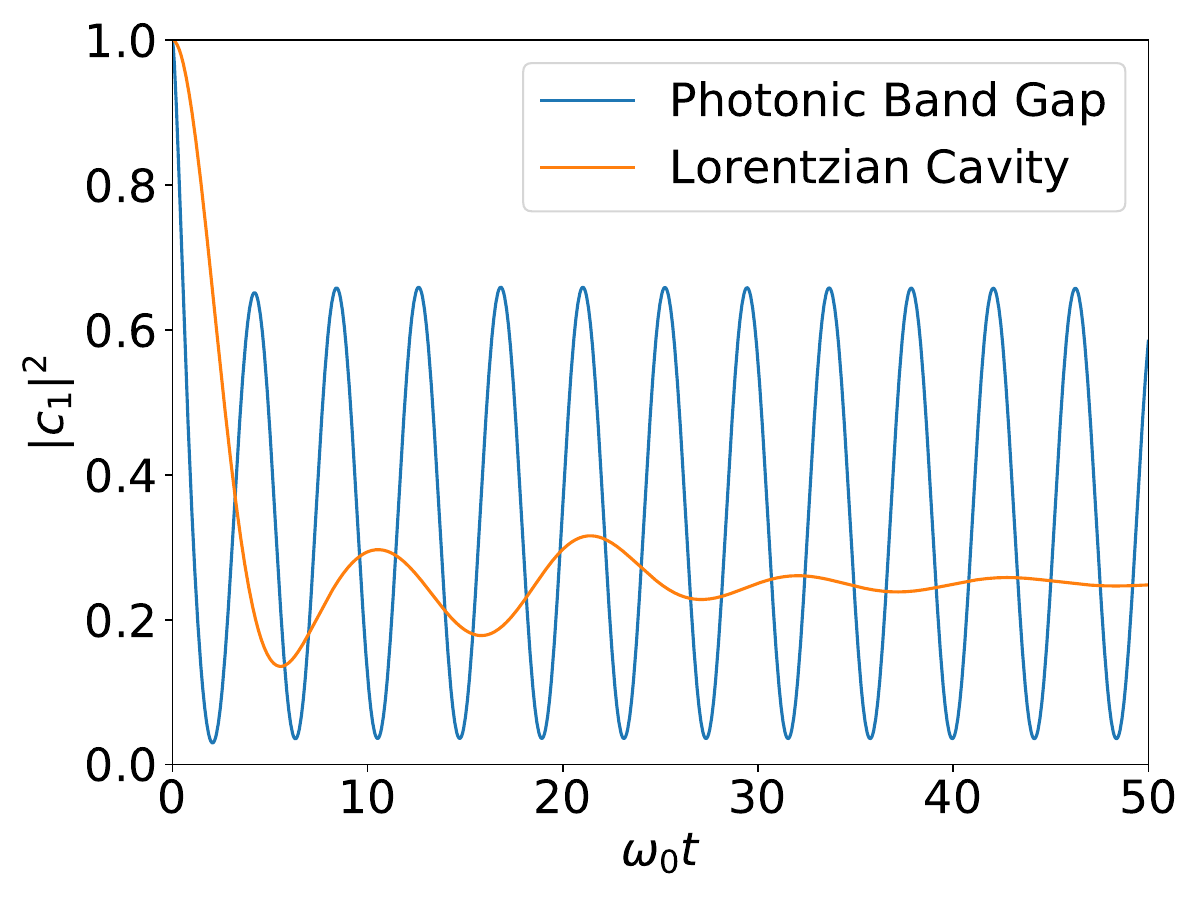}
         \caption{}
         \label{fig:PCvLC}
     \end{subfigure}
     \begin{subfigure}[b]{0.38\textwidth}
         \centering
         \includegraphics[height=0.8\textwidth]{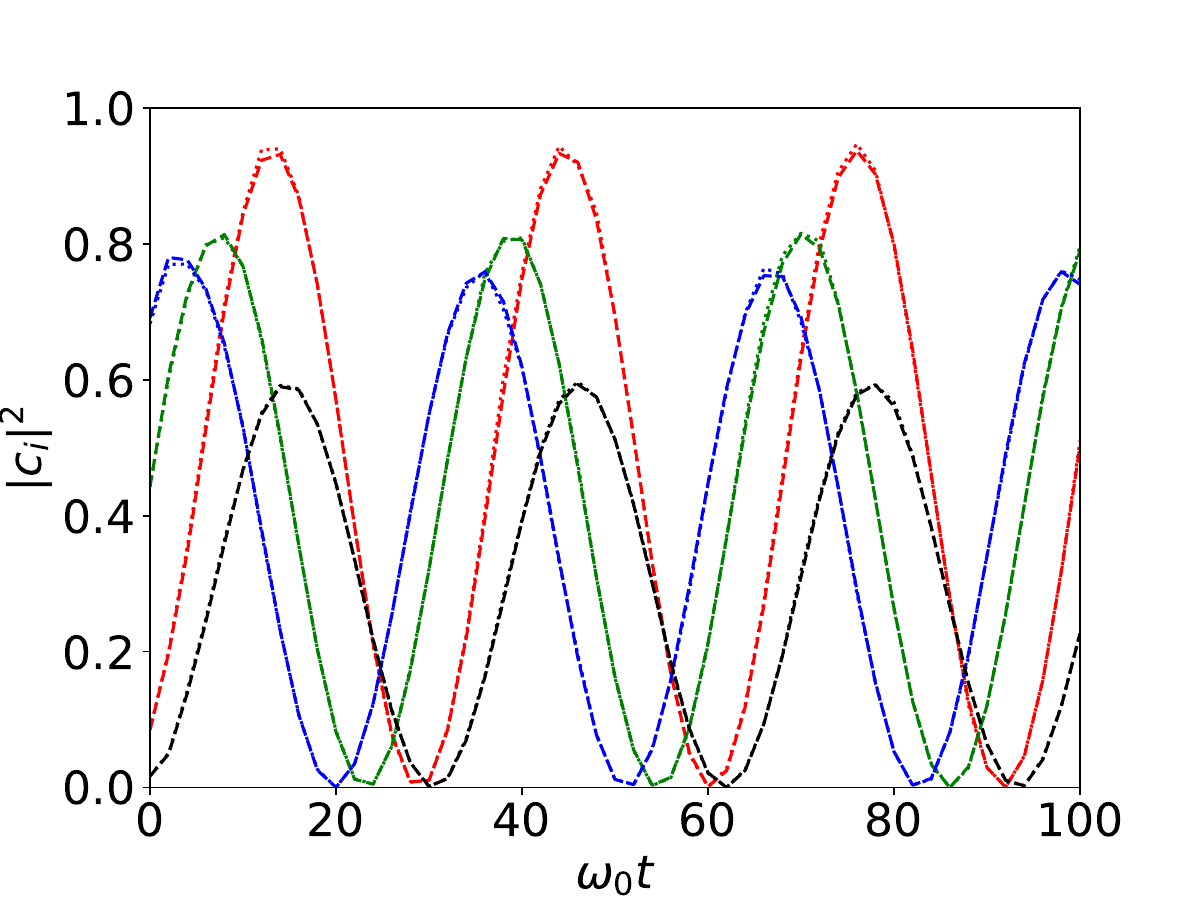}
         \caption{}
         \label{fig:QPred}
     \end{subfigure}
    \\
    \begin{subfigure}{0.45\textwidth}
         \centering
         \includegraphics[width=\textwidth]{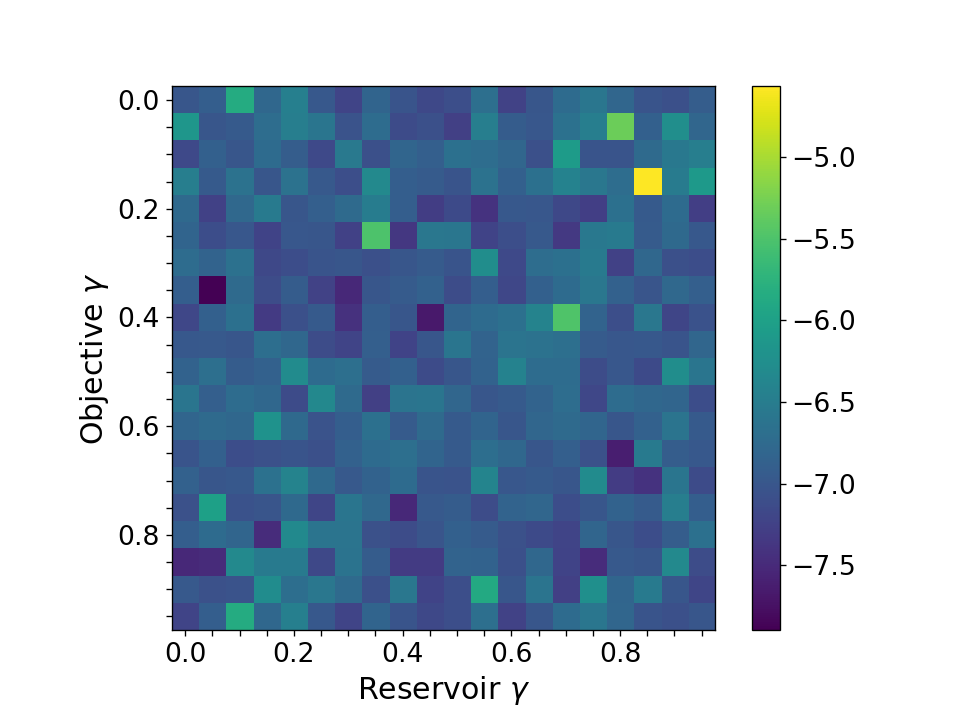}
         \caption{}
         \label{fig:QPredLCTOLC}
     \end{subfigure}
         \begin{subfigure}{0.45\textwidth}
         \centering
         \includegraphics[width=\textwidth]{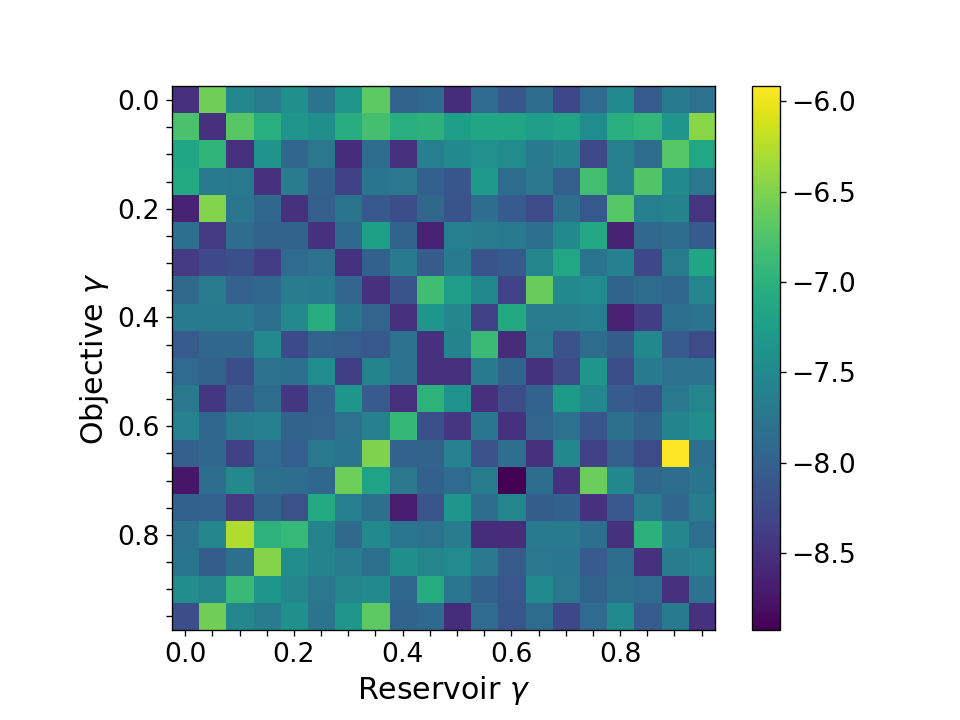}
         \caption{}
         \label{fig:QPredLCTOLCLong}
     \end{subfigure}
            \caption{(a) Schematic of the QPRC network. (b) Population dynamics of a single atom coupled to another atom inside of a photonic crystal (blue) and inside of a Lorentzian cavity (orange). (c) The QPRC prediction (dotted) and  exact (dashed) dynamics of two coupled atoms inside a photonic band gap material for three random initial conditions. The QPRC very accurately reproduces the exact dynamics for each initial condition.  (d,e) The heatmap of the average $\log_{10}$ error for QPRCs results compared to exact dynamics, where we utilise a Lorentzian cavity with spectral width given by the reservoir $\gamma$ to capture the dynamics of the objective Lorentzian cavity that has spectral width Objective $\gamma$ for two simulated run times: (d) $10~\omega_0^{-1}$ (e) $100~\omega_0^{-1}$ for varying values of both the reservoir and objective system's spectral width $\gamma$. }
        \label{fig:Q2Q}
\end{figure}

Up to now we have demonstrated that the QPRC approach can enhance the performance of conventional neural networks for pattern recognition tasks.  Here, we focus on employing  this approach to tackle quantum mechanical problems. Specifically, we explore the efficiency of the QPRC approach in determining the dynamics of open quantum systems, as  schematically presented in Fig. \ref{fig:Q2Qarch}. Most often, it is impractical if not impossible to fully describe the dynamics of the combined quantum system of interest and environment in open quantum systems. Various approximations are imposed to model the influence of the environment's degrees of freedom on the system of interest evolution. However, these approximations can have significant limitations. For example, a standard approximation in the field is the Markov approximation, wherein one assumes a unidirectional dissipation of information from the system into the environment. This approximation is only valid when the environment relaxes on timescales much faster than the systems such that the system barely perturbs the environment, akin to a weak coupling limit assumption, and fails rapidly for opens systems in which the environmental degrees display finite-bandwidth, long correlation times, or strong spectral features (e.g., quantum emitters strongly coupled to photonic crystal, waveguide, or cavity radiation reservoirs; spin boson models in solid-state quantum optics; atoms in optical lattices with dissipative coupling~\cite{Jin2014,Hood2016,Yu2019,PhysRevX.12.021042,PhysRevA.104.012213,PhysRevA.95.013626}).

The Markovian approximation has been successful in analyzing many systems, however, it falls short in capturing quantum-induced memory effects. For instance, in highly structured photonic systems, such as photonic crystals, the local density of states for the electromagnetic field undergoes rapid fluctuations for frequencies close to the edges of the photonic band gap. When a two-level atomic system is embedded with its transition energy close to the band edge of a photonic bandgap, the Markov approximation is rendered invalid due to these strong fluctuations~\cite{burgess_modelling_2021,SingleAtomSwitch,john_florescu_2001}, and extremely non-Markovian dynamics are exhibited. As a result of the strong interaction between the atomic system and the photonic reservoir, the atomic states become highly intertwined with the properties of the photonic reservoir and the degrees of freedom of the atoms. This results in temporal oscillations, fractional atomic population decay, spectral splitting,  sub-natural line widths of atomic transitions~\cite{John1994}, and enhancement of the coherence lifetime of multi-qubit systems by non-Markovian noise~\cite{White2020}.

Understanding the non-Markovian effects and quantum control processes associated with them may play a crucial role in the design and realisation of more reliable and performant quantum technologies, including quantum clocks ~\cite{QuantumNetworkClocks} and quantum networks~\cite{KimbleQuantumNetwork}. Beyond the preservation of entanglement, another challenge faced by quantum technology is the creation of such entanglement. An intriguing recent discovery demonstrates the viability of using the environment's non-Markovianity to create entanglement between atomic systems that were previously uncorrelated~\cite{EntangleGenNM,Fleming_2012}. 
We note that various methods exist for analyzing the dynamics of systems characterized by weaker non-Markovianity, such as memory effects arising from a smooth cut-off in the spectral density~\cite{RNNNonMarkov}. However, for quantum systems of considerably greater complexity and stronger non-Markovian behavior, such as those influenced by structured photonic reservoirs (including photonic crystals, waveguides, and cavities), new paradigms are necessary to effectively model these open quantum systems. This is due to the fact that non-Markovian effects can be generated by strong coupling to specific degrees of freedom within the environment, causing perturbative approaches to breakdown~\cite{pollock2013multi}. Furthermore, even in weakly coupled systems, if memory times are large for the environment, convolutionless perturbation theory also becomes incapable of accurately reproducing system dynamics~\cite{TheoryOQSBook}.

Here, we explore the ability  of a QPRC consisting of an atomic ensemble in a lossy cavity to capture the quantum dynamics of a strongly non-Markovian system. The system considered consists of a two-level atomic ensemble coupled to a photonic reservoir associated with a photonic band gap material. We are employing an isotropic one-dimensional photonic crystal band edge model for the photonic crystal, which can be directly implemented using photonic crystal waveguides~\cite{ResonanceFluorescence,burgess_modelling_2021}. This system generates strong atom-photon coupling, induced  by the square root divergence in the local density of states of the electromagnetic field modes around the band edge frequency  $\omega_I$ ($\rho(\omega)\propto(\omega-\omega_I)^{-\frac{1}{2}}$), which, in practice, can be considered to generate the highest degree of non-Markovianity in the atomic evolution. 

The Hamiltonian for the quantum system of interest is given in Eqn.~\ref{eqn:RWAH}, and leads  to the convolution of states given by Eqn.~\ref{eqn:ConvolutionRC}. However, due to the divergence in the density of states around the photonic band edge, the memory kernel takes the form 
\begin{equation}
   \Tilde{G}(s) = \beta^{3/2} e^{-i\pi/4}(s-i(\omega-\omega_I))^{-\frac{1}{2}},
\end{equation}
where $\beta$ is the coupling strength.
This divergence in the memory kernel leads to strong non-Markovian effects in the system generating a plethora of new features such as fractional decay of the atomic systems and photon-atom bound states ~\cite{John1994,john_florescu_2001,SingleAtomSwitch}. 

In Fig. \ref{fig:PCvLC}, we present the dynamics of a coupled pair of atoms inside a Lorentzian cavity and a photonic band gap material. Due to the specific structure of the system-environment coupling we can leverage symmetries in the Hamiltonian and deploy the rotating-wave approximation - removing highly oscillatory dynamics - to be able to calculate the non-Markovian dynamics from a fully Schr\"{o}dinger picture~\cite{burgessNatom,BurgessMemoryEffects}.  
We note that these two systems undergo very different dynamics. In the single atom case, the lossy cavity will dissipate all of the energy in the excited state at the steady state. In contrast, the atom placed inside the photonic band gap materials displays strongly non-Markovian dynamics. Due to the splitting of the energetic states by strong coupling to the lower photonic band edge some of the excitation is maintained even to the steady state giving rise to the fractional decay. 

Next, we train the QPRC atom-cavity network to model the non-Markovian dynamics of a pair of atoms in the photonic band gap system. To achieve this, we initialise the QPRC system with the same initial conditions (the input $c(0)$ coefficients)  as the photonic band gap system to be modelled. The output of the QPRC is then fed into a small neural network of 3 dense layers, which are subsequently updated via standard back-propagation. The results presented in Fig. \ref{fig:QPred} demonstrate the ability of atom-cavity QPRC to predict the dynamics of two atoms bound within the photonic band gap. Even for a relatively small sample size of 1500 data-sets and despite the distinctive dynamics of the two systems, the QPRC framework is very accurate in predicting the complex dynamics of the system.

Finally, we tested the efficacy of cavities to recreate the dynamics of cavities with different spectral widths. Here we have the same setup as in the previous experiment, with a single atomic system in both the target and QPRC system. To test multiple configurations of both the target and QPRC cavity systems, we varied the spectral width $\gamma$ of both. For each value of the QPRC cavity's spectral width we use the QPRC procedure to try to recreate the dynamics of the target system at each spectral width $\gamma$. For equal values of $\gamma$ the target and QPRC systems coincide.
The results of these numerical experiments are shown in Fig. \ref{fig:QPredLCTOLC} and demonstrate that the QPRC system is highly effective at capturing the dynamics of cavity systems even when the target and QPRC cavities have largely different spectral properties, showing average errors of $10^{-7}$ across the data sets. 

Furthermore, we extended the run time of the reservoir from $10\omega_0 ^{-1}$ to $100\omega_0^{-1}$ (see Fig. \ref{fig:QPredLCTOLCLong}), and remarkably we find that the efficacy of the reservoir increases greatly now with average errors in the range of $10^{-8}$. Some structure does appear in these errors. Notably, that the QPRC is best at reproducing the dynamics of a cavity that is the same as it's own depicted by the diagonal from the top left to the bottom right. The QPRC framework is not simply maintaining a representation of the input state but is performing useful computation due to the expansion of the input into a higher dimensional computational space brought about by its intricate non-Markovian dynamics.

\section*{Conclusion}
In summary, we have introduced a QPRC system consisting of two-level atomic systems coupled to Lorentzian photonic cavities as a model open quantum system. We applied the QPRC approach to an image recognition task using the MNIST database and compared the efficacy of the QPRC with a traditional neural network of the same design but without a QPRC layer. Surprisingly, our results showed that the QPRC initially underperformed compared to the traditional neural network when using relatively small training data sets. However, as the data set size increased, the QPRC rapidly surpassed the traditional neural network's performance. This finding is significant because modern neural networks often face limitations due to the size of the data they can effectively model. 
Additionally, we assessed the efficacy of the QPRC approach compared to the traditional neural network for different numbers of training epochs using a fixed data set size. Remarkably again, the QPRC consistently outperformed the traditional neural network at each epoch number examined. 
More importantly, we demonstrated the efficiency of the QPRC approach in capturing the dynamics of atomic systems coupled to cavities of various qualities, atomic systems in photonic bandgap materials, and other open quantum systems challenges. The QPRC proved highly effective in accurately representing the complex quantum dynamics of these systems, even with limited training data. The physical platform we introduced for QPRC opens up possibilities for a scalable tabletop quantum machine learning approach, leveraging experimentally verified quantum tomography methods~\cite{Lanyon2017}. This approach has the potential to significantly outperform conventional recurrent neural network approaches in various applications.

\printbibliography[heading=bibliography]
\end{refsection}

\section*{Data availability}
Source data for all figures are available on the Figshare data repository (https://doi.org/...). All other data that support  other findings of this study are available from the corresponding author upon reasonable request.

\bigskip



\noindent \textbf{\large Author contributions} 

\noindent A.B. initiated the project, performed simulations and wrote the paper.  M.F. initiated the programme, oversaw and directed the project and wrote the paper.

\bigskip

\noindent \noindent \textbf{\large Funding} 

\noindent This work was supported by the Leverhulme Quantum Biology Doctoral Training Centre at the University of Surrey funded by the Leverhulme Trust Training Centre under Grant No. DS-2017–079. M.F. acknowledges EPSRC (United Kingdom) under Strategic Equipment Grant No. EP/L02263X/1 (EP/M008576/1), EPSRC (United Kingdom) under Grant No. EP/016440/1 and EP/M027791/1 awards.

\bigskip 

\noindent \textbf{\large Competing interests} 

\noindent The authors declare no competing interests.

\bigskip 







\nocite{*}

 \end{document}